\documentclass{article}

\usepackage{forest}
\usepackage{tikz-qtree}
\usepackage{tikz}
\usetikzlibrary{shapes.geometric}
\usetikzlibrary{arrows.meta,arrows}
\usepackage{setspace}
\usepackage{arxiv}
\usepackage{hyperref}
\hypersetup{
colorlinks=true,
urlcolor=blue,
citecolor=blue}
\usepackage[utf8]{inputenc} 
\usepackage[T1]{fontenc}    
\usepackage{hyperref}       
\usepackage{url}            
\usepackage{booktabs}       
\usepackage{amsfonts}       
\usepackage{nicefrac}       
\usepackage{microtype}      
\usepackage{lipsum}
\usepackage{multirow}
\usepackage{adjustbox}
\usepackage{caption}
\usepackage{tabu}
\usepackage{tikz}
\usetikzlibrary{arrows}
\usepackage{smartdiagram}
\usesmartdiagramlibrary{additions}

\usepackage{stfloats}
\usepackage{lipsum}
\usetikzlibrary{trees}

\usepackage{graphicx} 
\graphicspath{{images/}} 

\title{An Overview of Collision Avoidance Approaches and Network Architecture of Unmanned Aerial Vehicles (UAVs)}

\author{
  Ahmad H. Sawalmeh\thanks{Correspondence: ahmad.sawalmeh@nbu.edu.sa}, ~~Noor Shamsiah Othman\\
  [0.8em]
  Department of Electrical and Electronics Engineering, Universiti Tenaga Nasional, 43000 Selangor, Malaysia \\
  [0.4em]
{PE20656@utn.edu.my,~Shamsiah@uniten.edu.my}\\
}

\begin{document}
\maketitle

\begin{abstract}
As an autonomous vehicles, Unmanned Aerial Vehicles (UAVs) are subjected to several challenges. One of the challenges is for UAV to be able to avoid collision. Many collision avoidance methods have been proposed to address this issue. Furthermore, in a multi-UAV system, it is also important to address communication issue among UAVs for cooperation and collaboration. This issue can be addressed by setting up an ad-hoc network among UAVs. There is also a need to consider the challenges in the deployment of UAVs, as well as, in the development of collision avoidance methods and the establishment of communication for cooperation and collaboration in a multi-UAV system. In this paper, we present general challenges in the deployment of UAV and comparison of UAV communication services based on its operating frequency. We also present major collision avoidance approaches, and specifically discuss collision avoidance approaches that are suitable for indoor applications. We also present the Flying Ad-hoc Networks (FANET) network architecture, communication and routing protocols for each Open System Interconnection (OSI) communication layers.
\end{abstract}

\keywords{UAVs \and FANETs \and Collision Avoidance \and  UAVs Communication Protocols}

\onehalfspacing

\section{Introduction}
Unmanned Aerial Vehicles (UAVs) is an emerging technology which has found its application in military \cite{lyon2004military,orfanus2016self}, as well as, public and civil applications \cite{cheng2014design,gupta2013review,hayat2016survey}. It has been widely used in military applications for more than $25$ years, which primarily consists of border surveillance, reconnaissance, and strike. Recently, there has been a growing interest of UAVs employment in many civilian applications, and its use will grow rapidly in the near future, UAV can be used in many applications. More specifically, there has been research on the employment of UAVs in many applications, such as: public safety, search and rescue, emergency communications, unexpected events, transportation management, remote sensing, scientific data collection, and industrial inspections, delivery of goods, precision agriculture. Finally, UAVs can also provide wireless coverage for ground users without using infrastructure communication systems, especially using UAVs as an aerial base station \cite{shakhatreh2018unmanned,shakhatreh2017efficient,chandrasekharan2016designing,shakhatreh2017maximizing,sawalmeh2017providing}.

Collision avoidance is one of the fundamental elements in an autonomous vehicles.
In \cite{pham2015survey}, the authors presented collision avoidance system in terms of sensing and detection, collision avoidance approaches and other factors. Moreover, the authors presented major categories of collision avoidance approaches, namely, geometric, optimized trajectory or path planning, bearing angle based approach (vision based) and force field. 
The authors in \cite{albaker2009survey}, summarized major collision avoidance approaches in the context of design factors, as well as, the advantages and disadvantages for each approach. More specifically, the authors categorized the major collision avoidance approaches into four main approaches, namely, predefined collision avoidance, protocol based decentralized collision avoidance, optimized escape trajectory approaches and E-Field approaches.

Table \ref{tableint1} delineates the closely related review papers on collision avoidance approaches and demonstrate the contributions of our review.

\color{black}
\begin{table*}
\footnotesize
	\renewcommand{\arraystretch}{1}
	\caption{\uppercase{COMPARISON OF RELATED WORKS ON COLLISION AVOIDANCE APPROACHES}}
	\label{tableint1}
	\centering
	\begin{tabular}{|c|c|c|c|c|c|c|c|}
		\hline 
		\multirow{ 2}{*}{Reference}&\multicolumn{3}{|c|}{Geometric Approaches} &{Potential} &{Path  } & {Vision}&{Indoor} \\
		\cline{2-4}
		&Point of Closest &Collision Cone &Dubins paths &Field & Planning& Based & Collision \\
		\hline
		\cite{pham2015survey}	&\checkmark&&&\checkmark&\checkmark&\checkmark&\\
		\hline
		\cite{albaker2009survey}	&\checkmark&&&\checkmark&\checkmark&\checkmark&\\
		\hline
		\cite{gardiner2011collision}	&\checkmark&\checkmark&\checkmark&\checkmark&\checkmark&\checkmark&\\
		\hline
		\cite{alexopoulos2013comparative}	&\checkmark&&&\checkmark&\checkmark&&\\
		\hline
		\cite{sedaghat2017collision}	&&&&&&\checkmark&\checkmark\\
		\hline
		This work&\checkmark&\checkmark&\checkmark&\checkmark&\checkmark&\checkmark&\checkmark\\
		\hline
	\end{tabular}
\end{table*}

On the other hand, setting up an ad-hoc network among UAVs is one of the most important design issues must be taken into account in multi-UAV systems. More specifically, communication between the UAVs is crucial for cooperation and collaboration in multi-UAV systems. This requires ad-hoc network between UAVs, which is referred to as Flying Ad-hoc Network (FANET).  Many researchers utilize the existing Mobile Ad-hoc Network (MANET) communication and routing protocols in FANET.
In \cite{bekmezci2013flying}, the authors presented the design characteristics of UAVs network. 
Then they discussed and summarized the main communication protocols for this network. They also introduced an open research issues for each protocol.
In \cite{sahingoz2014networking}, the routing protocols for FANET were presented based on the Open System Interconnection (OSI) communication layers, namely, physical layer, MAC layer, network layer, transport layer and cross-layer architectures. 
They categorized the routing protocols into four main classes, namely, static protocols, proactive protocols, reactive protocols and hybrid protocols. Then they presented FANET designed challenges, issues related to the existing FANET protocols, and future insights in FANET.
Tables \ref{tableint2} and \ref{tableint3} delineate the closely related surveys on FANET networking architectures and FANET routing protocols demonstrate the novelty of our review.

\color{black}
\begin{table*}
\footnotesize
	\renewcommand{\arraystretch}{1}
	\caption{\uppercase{COMPARISON OF RELATED WORKS ON FANET NETWORKING }}
	\label{tableint2}
	\centering
	\begin{tabular}{|c|c|c|c|c|c|c|c|c|}
		\hline 
		Reference&FANETs  &Data links and &Physical  &MAC  &Network  &Transport  &Cross-layer& Application  \\
		&Architectures&Operating Frequencies&Layer&Layer&Layer&Layer&&Layer\\
		\hline
		\cite{bekmezci2013flying}	&\checkmark&&\checkmark&\checkmark&\checkmark&\checkmark&\checkmark&\\
		\hline
		\cite{sahingoz2014networking}	&\checkmark&&&&\checkmark&&&\\
		\hline
		\cite{gupta2016survey}	&\checkmark&&\checkmark&\checkmark&\checkmark&&&\\
		\hline
		This work&\checkmark&\checkmark&\checkmark&\checkmark&\checkmark&\checkmark&&\checkmark\\
		\hline
	\end{tabular}
\end{table*}

\begin{table*}[!h]
\footnotesize
	\renewcommand{\arraystretch}{1}
	\caption{\uppercase{COMPARISON OF RELATED WORKS ON FANET ROUTING PROTOCOLS }}
	\label{tableint3}
	\centering
	\begin{tabular}{|c|c|c|c|c|c|c|}
		\hline 
		Reference&Static   &Proactive &Reactive   &Hybrid   &Geographic Based  &Hierarchical
		
		\\
		&Protocols&Protocols&Protocols&Protocols&Protocols&Protocols\\
		\hline
		\cite{bekmezci2013flying}	&\checkmark&\checkmark&\checkmark&&\checkmark&\checkmark\\
		\hline
		\cite{sahingoz2014networking}	&\checkmark&\checkmark&\checkmark&\checkmark&&\\
		\hline
		\cite{gupta2016survey}	&\checkmark&\checkmark&\checkmark&\checkmark&\checkmark&\\
		\hline
		This work&\checkmark&\checkmark&\checkmark&\checkmark&\checkmark&\checkmark\\
		\hline
	\end{tabular}
\end{table*}

This paper provides a survey on the major collision avoidance approaches, as well as, specifically discusses the main indoor collision avoidance techniques. The important element in cooperation and collaboration in multi-UAV system is addressed by presenting the FANET network architecture, communication and routing protocols for each OSI layer, as well as, specific protocols for application layer. 

This paper also highlights general challenges in the employment of UAVs, as  well as, main challenges specifically related to the development of collision avoidance techniques, as well as the establishment of communication for cooperation and collaboration of a multi-UAV system. Among the challenges are energy limitations, as well as regulation issues.

It is also important to understand different types of UAV communication services, specifically when considering UAV collision avoidance system and communication protocols in a multi-UAV system.

The rest of the paper is organized as follows. First in Section ~\ref{chall}, we present the main challenges in the employment of UAVs. Then in Section ~\ref{comm}, we discuss the UAVs communication aspects. In Section ~\ref{coll}, we present the main collision avoidance approaches. Next, Section ~\ref{FANET} presents the network architecture of the FANET. Finally, the conclusion is presented in Section ~\ref{con}.

\section{Challenges in UAV Deployment}
\label{chall}
\subsection{General challenges}
In the USA the number of UAVs is expected to increase from hundreds in $2015$ to over $230,000$ in $2035$ \cite{matolak2015unmanned}. This large number of UAVs could change the national airspace landscape, so safe integration of UAVs into the global airspace is an important concern in the deployment of UAV. One of the related issues in this situation is regulation issues, such as safe UAV operation, without breaking national security and public security. Technical issues, such as endurance time is an important issue to be addressed as the duration of UAV mission is lengthen. Whilst, UAVs autonomous operation encounters a critical issue, namely collision avoidance. Weather conditions is also a challenge that would affect the UAVs, due to its sensitivity to weather conditions and weather changes. For example, in a strong-wind condition, UAV may deviate from its predetermined path. The security aspect is an important challenge that need to be addressed. In \cite{javaid2012cyber}, the authors discussed the cyber security aspects of an UAV system. A detailed analysis of  possible threats to an UAV system was presented. With regard to communication issues, there is a need to address issue of providing a reliable communication, for example command and control links between UAV and the ground control station (GCS). In addition, a lack of radio spectrum protection can pose the possibility of lost of communication between a ground user and the UAV \cite{matolak2015unmanned,rosati2016dynamic}.

\subsection{Energy Consumption}
\label{Energy}
\begin{figure}
	\centering
	\includegraphics[scale=0.875]{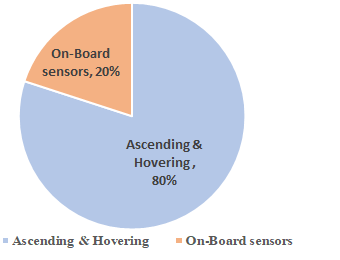}
	\caption{Energy consumption percentage for UAV}
	\label{fig:ahmad1-3}
\end{figure}

UAVs can be categorized based on energy source into four main types, fuel (kerosene) based UAVs, fuel cells, battery powered, and solar cells \cite{vergouw2016drone}. 
In general, UAVs consume energy to perform these tasks:(1) hovering and maneuver; (2) communication tasks, such as UAV-to-GCS, UAV-to-UAV, and UAV-to-Satellites; and (3) operation of on-board devices such as cameras and other sensors. 
There are two types of power sources for UAVs. In the first type, a single power source is used for all UAV equipments. More specifically, around 80\% of the energy is consumed for UAV ascending and hovering. Only 20\% is used by communication equipments and on-board sensors as depicted in Figure~\ref{fig:ahmad1-3}. 
In the second type, two power sources are used, one for ascending and hovering, and the other one for communication and on-board sensors.  
During UAVs missions, they need to maintain its power to complete the mission successfully. Therefore, it is important to find new techniques for conserving UAVs energy to increase UAVs flying time \cite{bekmezci2013flying}.

\subsection{Legislation}
A lack of the UAV regulations for the integration of UAVs flight into the  civilian airspace is one of the main issues that must be taken into consideration in the deployment of UAVs. Some national or civil aviation authorities has become the responsibility agency for the regulation and use of UAVs \cite{watts2012unmanned} in their respective countries.
As a matter of fact, the worldwide UAVs regulations are in its infancy. Therefore, the regulation issues of UAVs operations, such as safe UAVs operations without breaking national security, and public privacy became an important issue for governments \cite{stocker2017review}. 
In United states the Federal Aviation Administration (FAA) releases Federal Aviation Regulations (FARs) to regulate the UAV operations. They published guidelines, a policy document and a fact sheet, for the operation of UAVs below the $400$ feet, it is a must to apply for a Certificate of Authorization (COA) \cite{clarke2014regulation}. European Commission also introduced a rules of UAVs operations into European aviation safety system. The Law Library of Congress published a report that surveys the regulation of UAV operations under the laws of thirteen countries \cite{archick2005european}. On the other hand, many countries still do not have regulations governing the civilian uses of UAVs, and some of these countries do not allow to use UAVs in civilian applications. Therefore, it is important to legislate regulations and rules that govern civilian uses of UAVs. Moreover, this legislation must be enacted for a safe integration of UAV into the local and global airspace system.
\section{UAVs Communication}
\label{comm}
\begin{figure}[h]
	
	\centering\includegraphics[width=0.7\linewidth]{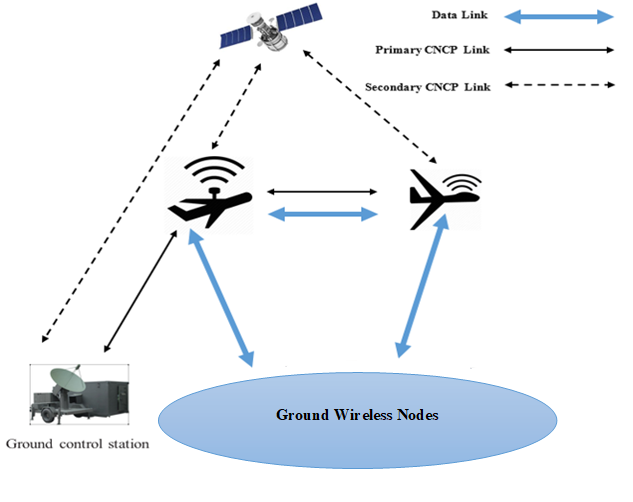}
	\caption{Types of UAV communication.}
	\label{UAV_comm}
\end{figure}

There are four main types of UAV communication services: (1) UAV-to-UAV for data and control links; (2) UAV-to-Ground Base Station for control and commands link; (3) UAV–to-Ground Wireless Nodes for UAV-aided data dissemination and collection; and (4) UAV-to-Satellite system \cite{zeng2016wireless}.
Figure ~\ref{UAV_comm} shows the wireless communication architecture for UAV networking. UAV wireless communication architecture consists of two basic types of communication links, namely the control link, which is referred to as Control and Non-Payload Communications (CNPC) Link, as well as, the data link.

\subsection{Control and Non-Payload Communications Link }
In general, CNPC is used for UAV safety, collision avoidance approach of sense-and-avoid information among UAVs. More specifically, CNCP provides more reliable and accurate information for UAVs operations. In CNCP link, control data are exchanged in the communication services of UAV-to-UAV, UAV-to-GCS, and GCS-to-UAVs. For safety purposes, low data rate is required for the exchange of control data in control link.

There are two types of CNCP available, namely, the primary CNPC link, which is the preferred control link, and the secondary CNPC link. The former link can be used via satellite as a backup link to enhance reliability and robustness. The primary CNPC link  is established directly during takeoff and landing. On the other hand the secondary CNPC link can be established via satellite when the UAV is in operation \cite{zeng2016wireless}. 

\subsection{Data Link}
Data link is used for sending and receiving data, namely, the downlink transmission from UAV to ground station or satellite, and uplink transmission from the ground station or satellite to UAV. 
In general, the capacity requirement of this data link depends on the applications \cite{zeng2016wireless}.

\subsubsection{Data Link Classification}
We classify the data link based on the operation frequency band, which are: (1) microwave band; (2) centimeter-wave (cmwave) band; and (3) millimeter-wave (mmwave) band.
Table \ref{freq} shows the comparison of these bands, in terms of frequency ranges, advantages and disadvantages, and pathloss models used on each band. 
The microwave spectrum is defined by ITU as a range of frequencies from $1$ to $6$ GHz. Microwave signal has large wavelength, which is able to penetrate into terrestrial infrastructures. The required transmission power for microwave signal is less than the higher bands, but the maximum rate of the data link in microwave band is $1$ Gbps. On the other hand, cmwave refers to the range of frequencies above $6$ to $33$ GHz. Thus, the required transmission power for cmwave signal is more than that of microwave signal, and has data rate better than microwave. Another frequency band is referred to as mmwave band. It can be used for high-speed wireless communications, which has frequencies span from $30$ to $300$ GHz, and the data rates can reach $10$ Gbps and more. However, it requires powerful transmitters, thus, more transmission power to cover the same coverage area than the lower frequency band. In addition, the mmwave band range requires Line-of-Sight (LOS), which cannot penetrate obstacles, and it is sensitive to environmental interference's.

\begin{table*}[!t]
	\footnotesize
	
	\renewcommand{\arraystretch}{1.7}
	\caption{\uppercase{Comparison of data link between different operation frequency bands}}
	\label{freq}
	\centering
	\begin{adjustbox}{width=1.0\textwidth}
		\begin{tabu} to 1.0\textwidth { | X[l] | X[l] | X[l] |X[l] | }
			\hline
			\textbf{Issue} &	\textbf{Microwave} & \textbf{cmwave} & \textbf{mmwave}\\
			\hline \hline
			Frequency Ranges &1-6 GHz &6-33 GHz&33-300 GHz \\ 
			\hline
			Frequency Bands &L, S, C &X, Ku, K&Ka, V, W, G \\ 
			\hline
			Advantages &1-Less transmit power; 2-Large ability to penetrate obstacles; 3-Low free space pathloss. &1-High data rate; 2-Better than microwave.  &1-Very high link data rate; 2-Used in high-speed wireless communications \\	
			\hline
			Disadvantages &1-Data link rate is limited.  &1-Free space pathloss more than microwave; 2-Need more power than microwave.
			& 1-High transmit power; 2-Sensitive to environmental interference; 3-High Free space pathloss
			\\
			\hline
			Path loss models &ATG, Outdoor-Indoor \cite{shakhatreh2018unmanned} &ATG, Outdoor-Indoor \cite{shakhatreh2018unmanned}&Outdoor only \cite{shakhatreh2018unmanned} \\
			\hline
			LOS, NLOS &LOS, NLOS &LOS, NLOS& LOS \\
			\hline
			Example Models  & \cite{al2014modeling,matolak2017air,feng2006path,data2003prediction}&\cite{imai2016outdoor}&\cite{samimi20153,akdeniz2014millimeter} \\	
			
			\hline

		\end{tabu}
		
	\end{adjustbox}
	
	\label{table:ahmad1-4}
\end{table*}

\subsubsection{Data Link Backhaul}
Backhaul is any point-to-point (P2P) communication link between remotely connected sites. In wireless communication, microwave and mmwave back-haul are becoming widespread. In data link, wireless microwave or mmwave backhaul is used for transmission of payload information between UAVs (UAV-UAV backhaul). In addition, it can also be used to allow UAV to communicate directly with the ground users (using UAV as an aerial base station).

\section{UAVs Collision Avoidance Approaches}
\label{coll}

An autonomous vehicles have common fundamental elements, such as ability of sensing and perceiving the environment, ability of analyzing, communicating, planning and decision making using on-board computers, as well as acting which requires vehicle contol algorithms. As there is no human control, UAVs are subject to collisions with obstacles either moving or stationary objects. This is one of the main autonomous UAV flight risks. Many collision avoidance methods have been proposed to address this challenge, and to solve the collision problem. The authors in \cite{pham2015survey} presented major categories of collision avoidance methods. However, there are more refined approaches to collision avoidance of the UAVs, which employs more sophisticated algorithm. More specifically, major collision avoidance approaches, namely geometric approach, potential field approach, path planning approach, and vision-based approach, as well as, indoor collision avoidance techniques, are presented in this section.

\subsection{Geometric Approach}
Geometric approach is a method used to avoid collision between two UAVs based on geometric equations, as well as, to decide UAVs direction in the conflict region \cite{park2008uav}. Moreover, this approach is used for path planning to avoid UAVs collisions with obstacles. 
There are different approaches to use this method to avoid collision that include point of closest approach (PCA) \cite{park2008uav}, Collision cone approach \cite{chakravarthy1998obstacle,mujumdar2009nonlinear}, and Dubins paths approach \cite{dubins1957curves,shanmugavel2010co}.

\begin{itemize}
	\item \textit{Point of Closest Approach (PCA)}: PCA is a collision avoidance method which is based on simple geometric approach. In this method, the UAVs are defined as a point mass with constant speed, and all UAVs in the geometric space are linked by database link, namely Automatic Dependent Surveillance-Broadcast (ADS-B) system to exchange the information, such as speed, altitude and GPS position among each other. In PCA a miss distance vector between two UAVs is calculated, and they are considered in a conflict region if this distance is less than the minimum allowed distance between UAVs \cite{park2008uav,gardiner2011collision}. 
	
	\item \textit{Collision Cone Approach}: The collision cone approach \cite{chakravarthy1998obstacle} can be used to predict collision possibilities between one UAV and other UAVs, and it can also be used to design collision avoidance method to avoid the collision. Here, UAV can avoid moving obstacles and/or other UAVs with unknown trajectories within the same geometric space as illustrated in Figure~\ref{fig:ahmad1}. 
	In this approach, a circle is placed around the obstacle/UAV that is to be avoided. Then a tangent line from this circle to the UAV is calculated. If there is one obstacle/UAV in between two tangent lines, then it is considered as critical.
	\cite{mujumdar2009nonlinear}
	\begin{figure}[!h]
		\centering
		\includegraphics[scale=0.45]{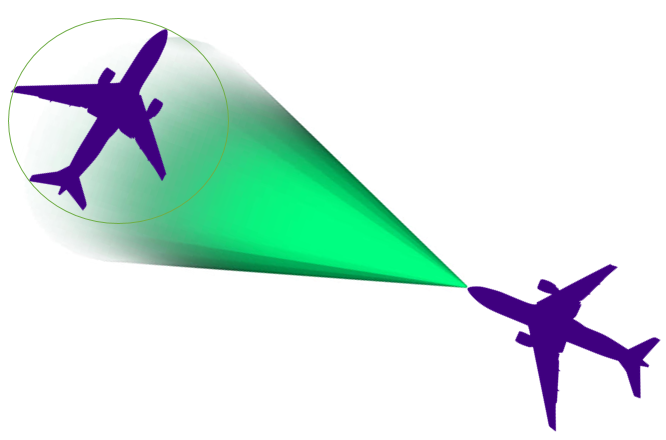}
		\caption{Collision cone Approach.}
		\label{fig:ahmad1}
	\end{figure}
	
	\item \textit{Dubins paths}: Dubins Path \cite{dubins1957curves} is another collision avoidance method that is based on geometry approach. It is the shortest path connecting two points in a plane with a constraint on  boundaries of the curvature or circular arc which is the shortest turn of the curvature \cite{gardiner2011collision}. A Dubins path is formed of three tangential circular arcs, or is formed by two tangential circular arcs with their common tangents in between.  In this method, UAVs need to compute trajectories that can help UAV to maneuver without collision with the obstacles \cite{shanmugavel2010co}.
	More Specifically, the Dubins path consists of five segments path that can be divided into initial, intermediate and final circular arcs, together with two straight lines in between. The initial and final circular arc are modified to make the line segments tangent to the obstacle circle \cite{tsourdos2010cooperative}. 
	However, This method assumes that the obstacles are stationary, so there is a need to extend to avoid moving obstacles. 
	
\end{itemize} 

\subsection{Potential Field Approach }
In \cite{khatib1986real}, the authors proposed this method to avoid the collision for ground robots. 
This method can also be used for avoiding a collision between UAVs and obstacles. In this method, the repulsive force fields are used which causes the UAV to be repelled by obstacles. The potential function is divided into attracting force field which pulls the UAV towards the goal, and repulsive force field which is assigned with the obstacle. UAV can follow collision free path, and does not collide with obstacles \cite{mujumdar2009nonlinear}. The potential field approach needs a large calculating power and time and this will make this method not suitable for small UAVs in real life applications \cite{pham2015survey}.


\subsection{Path Planning Approach}
\begin{figure}[!h]
	\centering
	\includegraphics[scale=0.4]{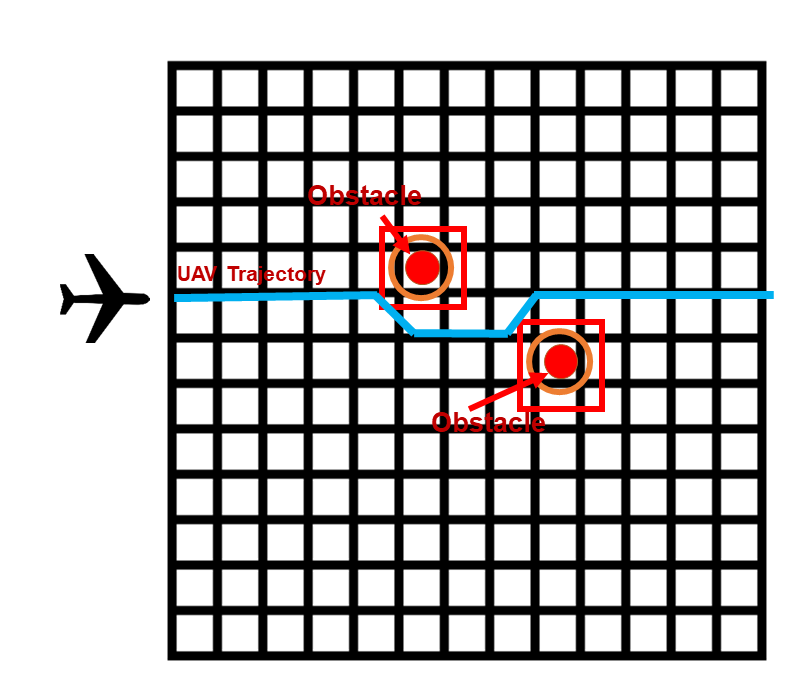}
	\caption{Path re-planning with A* search to avoid obstacles. }
	\label{fig:ahmad3}
\end{figure}

Path Planning is a grid based approach \cite{pham2015survey}, 
in this method, a collision free path can be found using path planning algorithms with graph search methods such as A*. This approach utilizes some geometric methods to calculate the collision free path during the flight. The flight map was divided into a grid as we shown in Figure~\ref{fig:ahmad3}
, then the collision free path is found using the search algorithm  \cite{alexopoulos2013comparative}.

\subsection{Vision-Based Approach}
\begin{figure}[!h]
	\centering
	\includegraphics[scale=0.35]{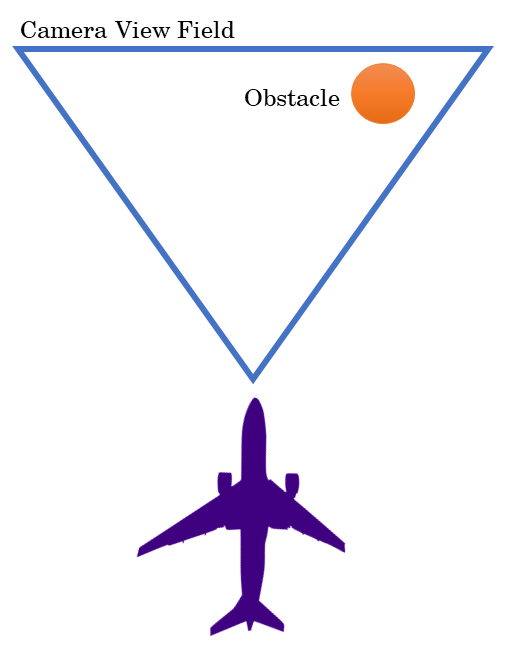}
	\caption{UAV approaching an obstacle, Vision-Based method to avoid obstacle collision. }
	\label{fig:ahmadn1}
\end{figure}

Recently, the vision-based obstacle detection using small cameras in UAVs becomes a topic of research interest. One of the limitations of this approach arise from the state of the sensors to reliably detect obstacles in large area. Moreover, most sensors are too heavy and cannot be used in UAVs. These sensors are normally used in large aircraft. This problem can be addressed with the advances in integrated circuits technology to design small and low power sensors and cameras that can be mounted on UAVs. This camera can help in observation, and obstacles collision avoidance using computer vision technology and algorithms. Moreover, the time to collision can be estimated using these cameras, and it can be provide object identification and object segmentation. This information UAV can be maneuvered around the obstacle without collisionas we can see in Figure~\ref{fig:ahmadn1} \cite{saunders2009obstacle}.

The main weakness of this method is the task complexity. Normally UAV is small and has a very limited resources. Therefore, it is not easy for an UAV to provide capabilities to see and avoid obstacles.
In \cite{sedaghat2017collision}, the authors utilized computer vision approach with UAV to avoid collisions with other UAVs. They used a machine learning algorithm referred to as cascade classifier to detect an UAV and camshift algorithm to track them. The coordinates of the center of an object moving towards the UAV, can be determined using using these algorithms. The distance between the UAV and the object can be calculated using stereo camera. If the obstacles is within less than 5 feet far from the UAV, a collision threat was sent, and avoidance maneuvers were executed. 
Nils Gageik et al \cite{7105819} proposed a simple solution for obstacle detection and collision avoidance of quadrotor UAVs by employing low cost ultrasonic and infrared range finders. This work proposed a solution which has low computational burden and did not require memory and time-consuming simultaneous localization and mapping.

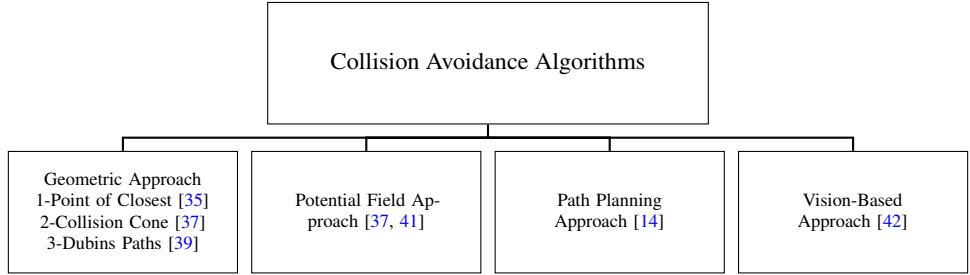
\begin{figure*}[!t] 
	\centering
	\begin{tikzpicture}[font=\scriptsize]
	\tikzset{every node/.style=
		{align=center, minimum height=46pt, text width=80pt}}
	\node[,draw=black] (b1) {Geometric Approach\\1-Point of Closest \cite{park2008uav}\\ 2-Collision Cone  \cite{mujumdar2009nonlinear}\\ 3-Dubins Paths \cite{shanmugavel2010co}};
	\node[right=5pt,draw=black] (b2) at (b1.east) {Potential Field Approach \cite{mujumdar2009nonlinear,khatib1986real}};
	\node[right=5pt,draw=black] (b3) at (b2.east) {Path Planning Approach \cite{alexopoulos2013comparative}};  
	\node[right=5pt,draw=black] (b4) at (b3.east) {Vision-Based Approach \cite{saunders2009obstacle}};
	\node[above=10pt, text width=160pt,draw=black] (top) at ($(b2.north)!.5!(b3.north)$) {\small{Collision Avoidance Algorithms}};
	\coordinate (atop) at ($(top.south) + (0,-5pt)$);
	\coordinate (btop) at ($(b3.south) + (0,-5pt)$);
	\draw[thick] (top.south) -- (atop)
	(b1.north) |- (atop) -| (b4.north)
	(b2.north) |- (atop) -| (b3.north);
	
	
	
	
	\end{tikzpicture}
	\caption{Main UAVs Collision Avoidance Methods}
	\label{fig:ahmad5}
\end{figure*}

In \cite{7535138} a vision-based collision avoidance approached was proposed that use the concept of dynamic safety envelope. In this work, the reliable communication link between the UAV and GCS is required. Image detection was processed by an onboard processor and the detection results will be transmitted to the GCS. The GCS executed the  proposed algorithm and transmitted commands for the avoidance maneuver back to the autopilot.
Figure~\ref{fig:ahmad5} summarizes the main UAVs collision and obstacle collision avoidance methods.

\subsection{Indoor Collision Avoidance}
The use of UAVs in the indoor environment is a challenging task and it needs higher requirements than outdoor environment. It is very difficult to use Global Positioning System (GPS) for avoiding collisions in the indoor environment, \textit{usually indoor is a GPS denied environment}. Moreover, RF signals cannot be used in this environment. RF signals could be reflected, and degraded by the indoor obstacles and walls. Therefore, several techniques can be utilized for indoor collision avoidance such as vision based methods using cameras and optical sensors, on-board sensors based method using IR, Ultrasonic and laser scanner, as well as, vision based combined with sensors based methods to provide accurate collision avoidance for indoor environments \cite{ luo2013uav}. 

The lightweight cameras on-board UAVs can be used efficiently for collision avoidance in indoor environments. It provides a real-time information about the walls and obstacles in this environment. Many researchers utilize this method to avoid collision and provide fully autonomous UAVs flight \cite{schmid2013stereo, alvarez2016collision, schmid2014autonomous, mustafah2012indoor}. The study in \cite{alvarez2016collision} used monocular camera with forward facing to generate collision free trajectory. Firstly, UAV took set of images to produce dense depth maps. Then, a suitable trajectory for navigation in forward direction was generated based on these maps. In this method, all computations were performed off-board and were sent in real time to computer over Wi-Fi network at about 30 frames per second, in order to perform image processing.

In \cite{mcfadyen2013aircraft}, the authors presented a reactive collision avoidance method for UAVs using spherical camera mode. This method uses vision-based collision avoidance techniques. It also used visual predictive and motion capture system to avoid obstacles. Here, the image processing operation was performed off-board. 

To solve the off-board image processing problem the authors in \cite{roelofsen2015reciprocal} proposed a collision avoidance system for UAVs with visual-based detection. The image processing operation was performed using two cameras, and small on-board computation unit. This method can be used to track several UAVs at the same time. This was due to the merging of derived measurements from the fish-eye cameras using color segmentation and Gaussian Mixture Probability Hypothesis Density (GM-PHD) filter. 

Vision-based collision avoidance methods using cameras, suffer from heavy computational operations for image processing. Moreover, these cameras and optical sensors are light sensitive, so steam and smoke could cause the collision avoidance system to fail. Therefore, sensor based collision avoidance methods can be used to tackle these problems \cite{chee2013control,gageik2012obstacle}. In \cite{ gageik2012obstacle} the authors proposed a sensor-based approach using ultrasonic sensors. This method contains two modules: obstacles detection module and collision avoidance module. In the obstacles detection module, 12 ultrasonic sensors with 360-degree circle were used to increase the obstacles detection accuracy. Moreover, in the collision avoidance module, the area around the UAV was divided into three main zones as ilustrated in Figure~\ref{fig:ahmad6}. These zones were:

\begin{figure}[!h]
	\centering
\includegraphics[scale=0.35]{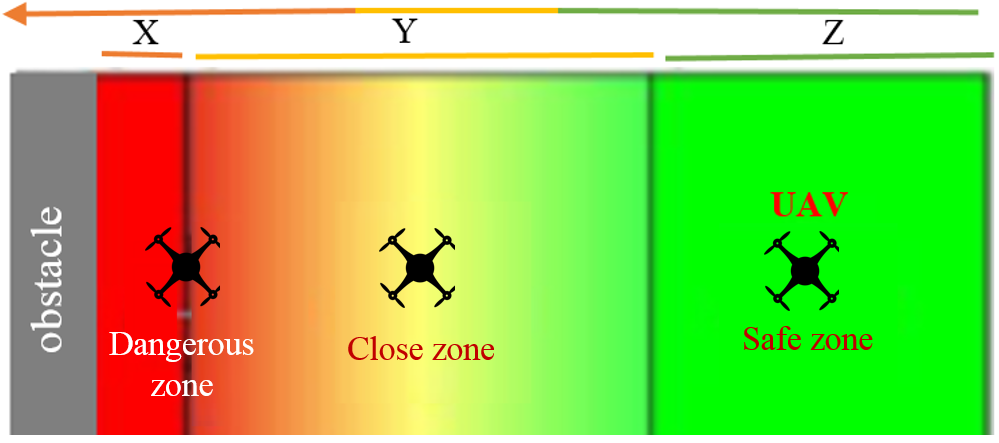}
	\caption{UAV collision zones. }
	\label{fig:ahmad6}
\end{figure}

\begin{enumerate}
	\item Green zone (safe zone), when UAV was far from any obstacles. In this zone the $obstacle\_distance > X+Y$, and the autonomous collision avoidance system was off.
	
	\item Yellow zone (close zone), when $X \textless obstacle\_distance \textless {X+Y}$. The close area state was activated and the angle and the approach speed was reduced.
	
	\item Red zone (dangerous zone), when obstacle was very close. More specifically, when $X > obstacle\_distance$, the dangerous zone was activated. Thus, the obstacle will be prevented from approaching the obstacle.   
\end{enumerate}

Vision based collision avoidance methods are light sensitive, so they need sufficient lighting to work properly. Moreover, sensors based methods require structured environments. For example, unorganized surfaces and far obstacles cannot be detected using ultrasonic sensors. Therefore, on-board cameras and exteroceptive sensors such as ultrasonic sensors, laser scanner, inertial sensor (IMU) can be combined for avoiding obstacles and to provide fully autonomous indoor/outdoor flights for UAVs \cite{ gageik2012obstacle}.

Many researchers utilized this approach to detect and avoid collisions for UAVs. The study in \cite{huh2013integrated} proposed an integrated navigation model using camera, gimbaled laser scanner and IMU for autonomous flight of UAVs for outdoor and indoor environments. They proposed to calibrate camera and laser sensor with each other using a simple visual marker. They also proposed a real time navigation system based on the Extended Kalman Filter (EKF) with simultaneous localization and mapping (SLAM) algorithm which was appropriate for this model.
In \cite{tomic2012toward} the researchers used two exteroceptive sensors, 2D laser range finder (LRF) and stereo cameras to enable six degrees of freedom (DoF) for vision obstacles detection and fully autonomous UAVs flights. This approach is restricted to the front field of view (FoV) of the stereo cameras. On the other hand, the study in \cite{chambers2011perception} used cameras and sensors together and it took into account obstacles in all directions.

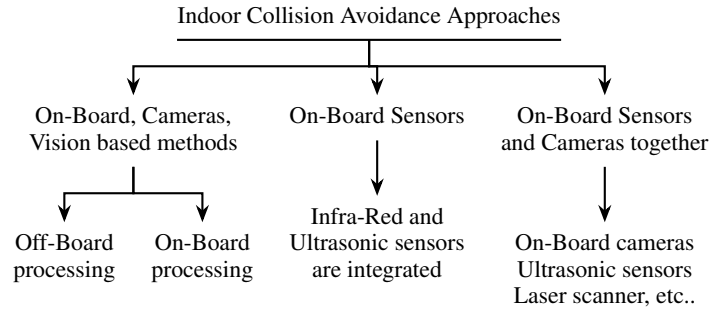
\begin{figure}
	\centering
	\begin{forest}
		for tree={
			align=center,
			parent anchor=south,
			child anchor=north,
			font=\footnotesize,
			edge={thick, -{Stealth[]}},
			l sep+=10pt,
			edge path={
				\noexpand\path [draw, \forestoption{edge}] (!u.parent anchor) -- +(0,-10pt) -| (.child anchor)\forestoption{edge label};
			},
			if level=0{
				inner xsep=0pt,
				tikz={\draw [thick] (.south east) -- (.south west);}
			}{}
		}
		[  Indoor Collision Avoidance Approaches 		
		[{On-Board, Cameras,\\ Vision based methods}
		[{{Off-Board\\ processing}}		
		]
		[{On-Board \\ processing}		
		]
		]
		[On-Board Sensors 
		[{Infra-Red and \\ Ultrasonic sensors\\ are integrated }
		]
		]
		[On-Board Sensors\\ and Cameras together
		[{On-Board cameras \\ Ultrasonic sensors\\Laser scanner, etc.. }
		]
		]
		]
	\end{forest}
	\caption{Indoor collision avoidance approaches for UAVs.}
	\label{fig:ahmad7}
\end{figure}

We classify the main indoor collision avoidance approaches for UAVs based on technology and types of processing as shown in Figure~\ref{fig:ahmad7}.

\section{FANETs Networking Architectures}
\label{FANET}

\begin{figure}[!h]
	\centering
	\includegraphics[scale=0.5]{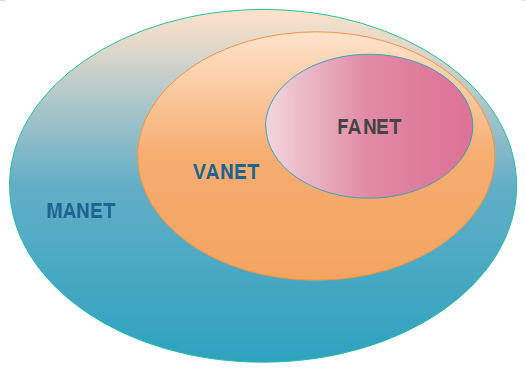}
	\caption{FANET subset of VANET, and both subgroub of MANET}
	\label{fig:ahmad8}
\end{figure}

Due to the high degree of mobility for UAV, MANETs and VANETs does not meet all requirements of the UAV networks. UAV network is dynamic, with nodes changing its positions due to a randomly continuous movement in two or three dimensions, so the UAV network is characterized by fluid topology \cite{gupta2016survey}.
Changes in UAV network topology is much more frequent than other networks. The routing protocol implementation for the UAVs network is an important concern. The MANET routing protocols may fail in tracking UAVs network changes. 
Moreover, UAVs have on-board energy limitations \textit{(usually UAV is battery powered)}, power consumption should be taken into account for routing protocol implementation of UAVs network.
Furthermore, in case of out-of-service for one of UAV swarm, the challenge arises on how to transfer user link to another active UAV seamlessly, without disruption of user session. In order to manage these challenges FANETs are used, which is an ad-hoc network connecting the UAVs. FANET basically is a special type of MANET. More specifically, VANET and FANET are a subset of MANET as we can see in Figure~\ref{fig:ahmad8} \cite{gupta2016survey}.
The network topology in FANET can change much more frequently than the other two network types MANET \cite{bekmezci2013flying}. 
We make a comparison between MANET \cite{bekmezci2013flying} and FANET \cite{gupta2016survey}, as presented in Table~\ref{table:ahmad1-4}.

\begin{table*}[!t]
	\scriptsize
	
	\renewcommand{\arraystretch}{1.7}
	\caption{Comparison between MANET and FANET}
	\label{table}
	\centering
	\begin{adjustbox}{width=1.0\textwidth}
		\begin{tabu} to 1.0\textwidth { | X[l] | X[l] | X[l] | }
			\hline 
			Issue & MANET& FANET \\
			\hline \hline
			Mobility model  & 
			1-Random mobility model.
			& 
			1-Predetermined path, mobility model is regular.    
			2-Not-Predetermined, (many mobility models \cite{bouachir2014mobility,kuiper2006mobility}), e.g. Semi Random Circular Movement (SRCM).
			
			\\
			\hline
			Topology & Ad-hoc Network & 1-Mesh between UAVs.
			2-Star with control station
			3- Ad-hoc
			
			\\
			\hline
			Topology changes & Dynamic topology but changes less than FANET (Infrequent) & Fluid topology, (Frequent)   
			
			\\
			
			\hline
			Radio propagation model & Close to ground, NLOS& Far away from the ground most cases, LOS. 
			
			\\
			\hline
			Power consumption  & Node is battery-powered& Fuel based UAVs, or battery powered, or solar cells 
			
			\\
			\hline
			Computational power  & nodes can act as routers, so it's must to have computation capabilities& UAV can use specific devices with high computational
			Power.
			
			\\
			\hline
			Node density (Number of nodes in a unit area)  & High node density& Low node density

			\\
			\hline
			Node speed & Low  &  High 
			
			\\
			\hline
			Security requirements & Medium  &  High 
			
			\\
			\hline
		\end{tabu}
		
	\end{adjustbox}
	
	\label{table:ahmad1-4}
\end{table*}

In single UAV system, a UAV needs to communicate with GCS, usually UAV in this system is large and need heavy communication equipments to establish a communication link with the GCS. Moreover, the cost of UAV failure is expensive. On the other hand, using small UAVs with multiple UAVs system has many advantages than the single UAV system. The coverage area will increase for multiple UAVs system. While in a single UAV the coverage area is limited. In addition, the cost of small UAVs are less than single large UAV. Furthermore, the use of multiple UAVs system minimizes the time required to complete the mission. Moreover, in single UAV system if UAV fails it must return to the GCS and the task cannot be completed. However, in multi UAV system the other UAVs can complete the task, thus, the fault tolerance will be increased for the system \cite{tareque2015routing}. In FANETs communications, OSI is usually used as a communication model for UAVs, which includes the physical, medium access control (data link), network, transport and application layers.    
\subsection{Physical Layer}
Physical layer deals with the hardware implementation, signal transmission technologies, path loss models, links and channel characteristics, connectivity to neighbors, encoding and signaling implementation, and moves the data bits over the physical medium using different types of antennas. 
For more details about physical aspects refer to Section ~\ref{comm} and \cite{shakhatreh2018unmanned}.

\subsection{Medium  Access Control Layer (MAC) } 
The MAC layer is responsible to handle physical layer errors, channel access policies, flow and error control. MAC layer uses the general IEEE 802.11 protocol \cite{ieee1999part11} and its extensions such as IEEE 802.11g/n \cite{gu2015airborne}, and usually omni-directional antenna is used for communication, which communication range can exceed several hundred meters in line of sight (LOS) communication \cite{jawhar2017communication}. 
In FANETs, there are several issues that need to be considered in designing efficient MAC layers protocols such as: (1) the high 3D movement of UAVs; (2) large distance between UAVs; (3) intermittent connectivity of networks; (4) frequent links changes and outages. Moreover, the designers need to guarantee low packets error rate and low latency to increase throughput for this layer. 
The traditional designs for MAC layer use half-duplex radios techniques and do not support multi-packet reception (MPR) capability \cite{ho2010qos}. In this kind of communication, the UAV node cannot receive packets while transmitting to avoid the collision between incoming packets. To tackle this problem, the authors in \cite{cai2013medium} proposed a MAC schemes for FANETs with token-based protocol, full-duplex radios and MPR. More specifically, in this work, full-duplex allowed UAV nodes to receive and transmit packets at the same time and MPR allowed UAV to receive concurrent packets.
Moreover, a directional antenna can be used to address some of these challenges and increase the communication range for FANETs. The authors in \cite{alshbatat2010adaptive} proposed adaptive MAC protocol for FANETs using directional antennas. They used omni-directional antenna for control packets Request To Send (RTS), Clear To Send (CTS) and Acknowledgment (ACK). Moreover, the directional antenna is used to send data packets. 
In \cite{temel2015lodmac}, the authors presented a new mac protocol for FANETs, referred to as  Location Oriented Directional MAC protocol (LODMAC). In this protocol, directional antennas and location estimation for neighbors nodes were utilized within the MAC layer. Moreover, along with CTS and RTS packets, it uses a busy to send (BTS) as a new control packet. LODMAC can address the deafness problem of directional MAC and it outperformed the other adaptive MAC protocol for FANETs using directional antennas.

\subsection{Network Layer} 
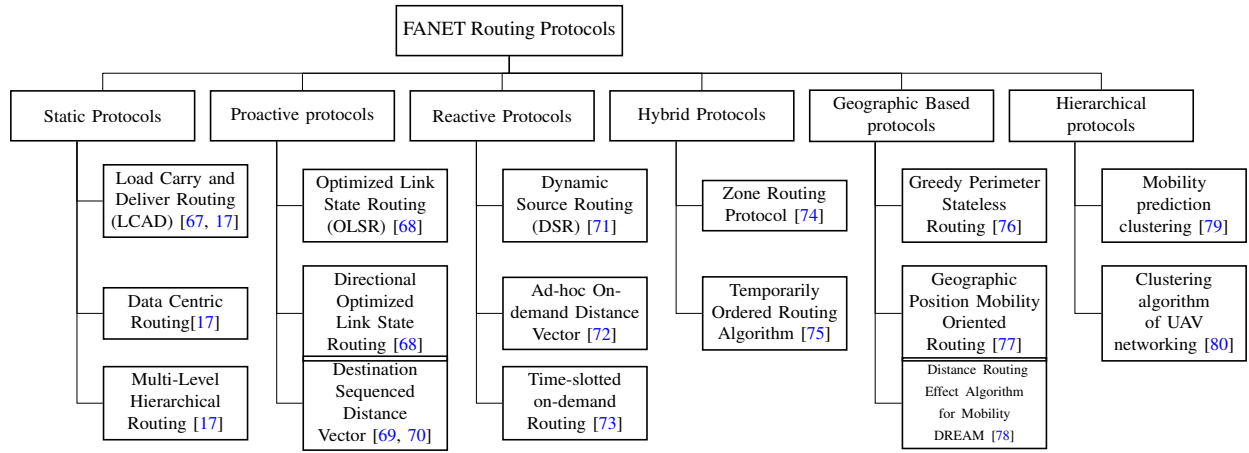
\begin{figure*}[!t] 
	\resizebox{\linewidth}{!}{
		\begin{tikzpicture}[
		Models/.style={text centered, text width=3cm},
		attribute/.style={%
			grow=down, xshift=0cm,
			text centered, text width=2.3cm,
			edge from parent path={(\tikzparentnode.225) |- (\tikzchildnode.west)}},
		first/.style    ={level distance=10ex},
		second/.style   ={level distance=22ex},
		third/.style    ={level distance=32ex},
		fourth/.style   ={level distance=40ex},
		fifth/.style    ={level distance=48ex},
		level 1/.style={sibling distance=10em}]
		\node[anchor=south]{FANET Routing Protocols}
		[edge from parent fork down]
		
		child{node (model1) [Models] {\footnotesize{Static Protocols}}
			child[attribute]  {node {\footnotesize{Load Carry and Deliver Routing (LCAD) \cite{cheng2007maximizing,sahingoz2014networking}}}}
			child[attribute,second] {node {\footnotesize{Data Centric Routing\cite{sahingoz2014networking}}}}
			child[attribute,third]  {node  {\footnotesize{Multi-Level\\ Hierarchical Routing \cite{sahingoz2014networking}}}}}
		child{node [Models] {\footnotesize{Proactive protocols}}
			child[attribute,first]  {node {\footnotesize{Optimized Link State Routing (OLSR) \cite{clausen2003optimized}}}}
			child[attribute,second] {node {\footnotesize{Directional Optimized Link State Routing \cite{clausen2003optimized}}}}
			child[attribute,third]  {node  {\footnotesize{Destination Sequenced Distance Vector \cite{perkins1994highly,alshabtat2010low}}}}}
		child{node [Models] {\footnotesize{Reactive Protocols}}
			child[attribute,first]  {node {\footnotesize{Dynamic Source Routing (DSR) \cite{johnson1996dynamic}}}}
			child[attribute,second] {node {\footnotesize{Ad-hoc On-demand Distance Vector \cite{murthy1996efficient}}}}
			child[attribute,third]  {node  {\footnotesize{Time-slotted on-demand Routing \cite{forsmann2007time}}}}
		}     
		child{node [Models] {\footnotesize{Hybrid Protocols}}
			child[attribute,first]  {node {\footnotesize{Zone Routing Protocol \cite{haas2002hybrid}}}}
			child[attribute,second]  {node {\footnotesize{Temporarily Ordered Routing Algorithm \cite{park2001temporally}}}}}
		child{node [Models] {\footnotesize{\footnotesize{Geographic Based\\ protocols}}}
			child[attribute,first]  {node {\footnotesize{Greedy Perimeter Stateless Routing \cite{shirani2011performance}} }}
			child[attribute,second]  {node {\footnotesize{Geographic Position Mobility Oriented Routing \cite{lin2012novel}}}}
			child[attribute,third]  {node  {\scriptsize{ Distance Routing Effect Algorithm for Mobility DREAM \cite{peters2011geographical}}}}
		}
		child{node [Models] {\footnotesize{Hierarchical \\ protocols}}
			child[attribute,first]  {node {\footnotesize{Mobility prediction clustering \cite{zang2011mobility}} }}
			child[attribute,second]  {node {\footnotesize{Clustering algorithm of UAV networking \cite{liu2008clustering}}}}};
		\end{tikzpicture}}
	\caption{Classification of FANET routing protocols}
	\label{fig:ahmad9}
\end{figure*}

In FANET, the initial experiments were implemented using the existing MANET routing protocols. MANET routing protocol can be classified into six main classes which are: static, proactive, reactive, hybrid, geographic based, and hierarchical protocols. MANET routing protocols do not satisfy all requirements of FANET network; Therefore, the existing MANET protocols were modified to be used in FANET network. We classify these protocols and provide examples for each protocol, as shown in Figure~\ref{fig:ahmad9}. 

\tikzstyle{every node}=[draw=black,thick,anchor=west, minimum height=2.5em]

\subsection*{\textbf{5.3.1. Static Protocols}}
In this protocol, the UAV routing information is loaded onto the UAV before its mission starts. During the mission, the routing table is fixed and it cannot be changed \cite{sahingoz2014networking,tareque2015routing}. More specifically, the UAV network topology and routing information cannot be changed until the UAV mission ends. Moreover, UAV needs to communicate with other UAVs in the same FANET coverage or with GCS/satellite. To update UAV routing tables, it needs to wait until the mission ends. Therefore, there is no fault tolerant in static routing protocol. The static protocols are:

\begin{itemize}
	\item {Load Carry and Deliver Routing (LCAD):}
	In this protocol, UAV is used to carry data from one ground station to another with single hop communication. The source ground station loads the data to UAV, and then UAV carries it towards the destination ground station as depicted in Figure~\ref{fig:ahmad10}$-a$. The main advantages of LCAD include the freeness of interference communication and this maximizes the throughput. Moreover, because of this model is single hop communication, it will be considered as a secure routing protocol. But the main drawback of this model is the delay of data delivery. To solve this problem, multiple UAVs system can be used between source and destination to decrease the time delay.
	\cite{cheng2007maximizing,sahingoz2014networking}. 
	\begin{figure}[!h]
		\centering
		\includegraphics[scale=0.32]{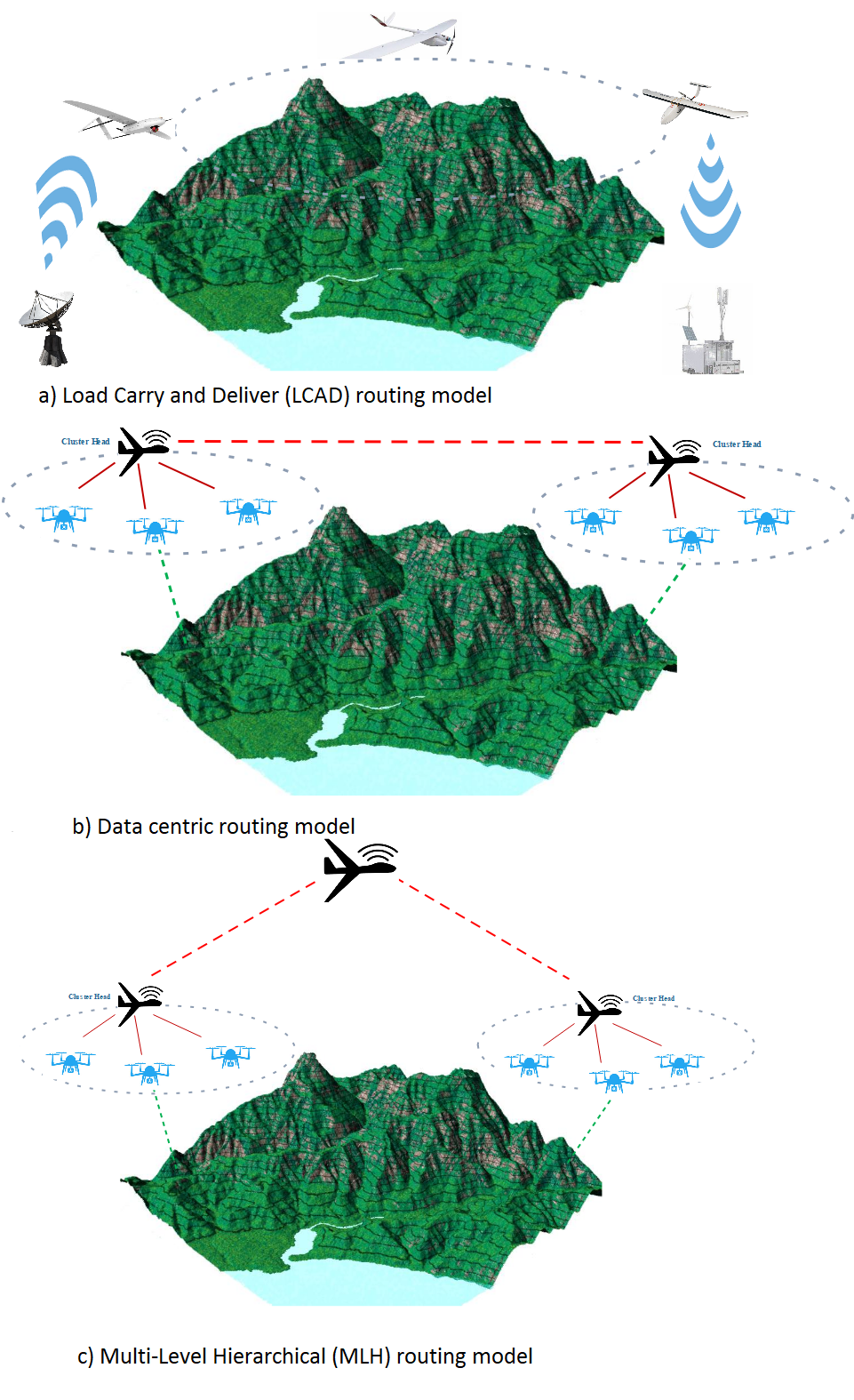}
		\caption{Static routing models $a)$LACD routing model. $b)$Data centric routing model. $c)$Multi-Level Hierarchical Routing (MLH) model}
		\label{fig:ahmad10}
	\end{figure}
	
	\item {Data Centric Routing (DCR):}
	Data centric routing is used when the data is requested by more than one UAV in FANET. This protocol is preferred in one to many communications in FANET, and the routing is done according to the data contents \cite{sahingoz2014networking}.  
	In this model, the query and request data dissemination will add some extra load to the network, due to the redundant data that is sent to the network \cite{bekmezci2013flying}.
	
	\item {Multi-Level Hierarchical Routing (MLH)}:
	In MLH, routing swarm of UAVs is organized into hierarchical model and the FANETs are divided into clusters of UAVs where, each cluster consists of cluster head and cluster members \cite{zang2011mobility}. This model can be used to increase the scalability and the mission area of the FANETs, This model is appropriate for large number of UAVs in FANETs \cite{sahingoz2014networking,bekmezci2013flying}.  
\end{itemize}

Due to the dynamic environment, fluid topology, and rapid changes in links between UAVs, static routing protocol is not widely used in FANET. Thus, dynamic routing protocol can be used to fulfill the FANET requirement. Dynamic routing protocols are divided into proactive or table driven protocol, and reactive or on demand routing protocol.  	
\subsection*{\textbf{5.3.2. Proactive Routing Protocols (PRP)}}
In this protocol, every node contains a table that is used to store all routing information about all other nodes in FANET. So, each node knows everything about other nodes, and it always has the latest routing information about other nodes in network. Therefore, sender node can select the path to the receiver node directly. On the other hand, in this kind of protocols, the bandwidth cannot be efficiently used because PRP needs to exchange a lot of messages between the nodes to refresh the routing tables. For this reason, PRP is not suitable for large FANET, or in high mobile networks, and can be used in small FANET only \cite{clausen2003optimized,perkins1994highly,alshabtat2010low}. 
\begin{itemize}
	\item {Optimized Link State Routing (OLSR)}
	is a table driven, proactive protocol designed to use in MANETs. Every node stores all routing information about all other nodes at start-up, and the node topology information will be exchanged with other nodes in the network periodically using two types of messages. Hello messages are sent for detecting the connectivity with neighbor nodes by broadcasting to one hop. Also, control messages are issued to keep topology information of the network updated. These messages issued periodically to refresh the topology information of network. The number of exchanged messages are very large in this protocol, this is due to the broadcasting information of every node to all other neighbor’s nodes, so the overhead of the network will increase. To tackle this challenge some nodes are used as a Multi-Point Relay (MPR) as depicted in Figure~\ref{fig:ahmad11}. These nodes are responsible to forward the control messages. The number of control messages to be exchanged can be reduced using this mechanism \cite{sahingoz2014networking,clausen2003optimized}. 
	\begin{figure}[!h]
		\centering
		\includegraphics[scale=0.45]{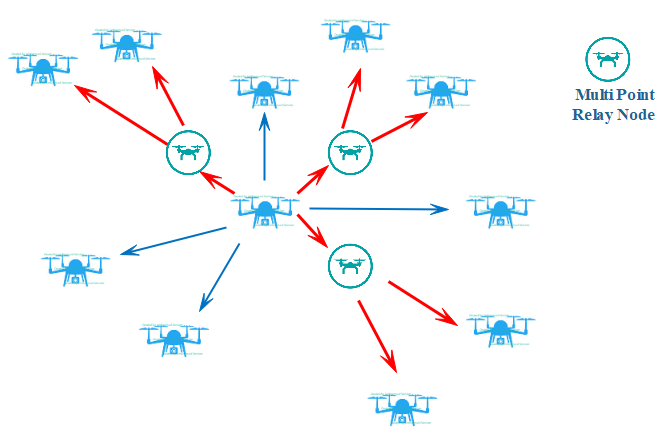}
		\caption{Multi-Point Relay (MPR) model }
		\label{fig:ahmad11}
	\end{figure}
	This protocol fails to provide reliable communications for UAVs network \cite{gupta2016survey}; therefor,  Directional Optimized Link State Routing Protocol (DOLSR) can be used to solve this problem.
	
	\item {Directional Optimized Link State Routing}
	DOLSR protocol is proposed in \cite{alshabtat2010low} and it is based on OLSR protocol \cite{clausen2003optimized} but with directional antenna. In this protocol, directional antennas and heuristic technique are used to minimize the number of MPR and to reduce the total number of control messages to be exchanged in the FANET. Apart from this, DOLSR can minimize the end-to-end network delay, where the sender node measures the distance to the destination,  if the distance is larger than half of the maximum distance, than it can be accomplished using directional antenna DOLSR which is used by the sender node. Otherwise, the sender node will use the OLSR \cite{alshabtat2010low}.   
	
	\item {Destination Sequenced Distance Vector (DSDV)}
	is a table driven, proactive routing protocol, bellman Ford algorithm was used with small modification to be suitable for FANET. In this protocol, every node saves the routing table for all other nodes, a sequence number generated by the destination node is assigned to each entry in the routing table to avoid the routing loop problem. At any changes in the network topology the protocol disseminates these changes to all other nodes in the network to keep their routing information up to date. Solving routing loop problem, and determining the freshness of a route are the main advantages of this protocol. On the other hand, this protocol needs large network bandwidth for updating the routing tables, this will cause overhead to the FANET.
	\cite{perkins1994highly,alshabtat2010low,he2002destination}.  
	
\end{itemize}

\subsection*{\textbf{5.3.3. Reactive Routing Protocols (RRP)}}
RRP is an on-demand bandwidth efficient routing protocol used for ad-hoc networks when there is no route between the sender and receiver nodes. In RRP, routes are discovered and maintained as needed, this will reduce the overhead occurred by proactive routing protocols. The sender node is responsible to initiate the route discovery process with the receiver node. Thus, the sender will flood the network with a Route-Request packet and after that the destination replies to the request with Route-Reply packet. This process will reduce the power consumption, save network bandwidth, and solve the routing overhead problem of proactive routing protocols. On the other hand, for every unreachable node route, search process is needed. This process takes a long time. Therefore, it will be slower than proactive protocols. In RRP, there are three key protocols: Dynamic Source Routing (DSR) \cite{johnson1996dynamic}, Ad-hoc On-demand Distance Vector \cite{murthy1996efficient}, and Time-slotted On-demand Routing \cite{forsmann2007time}.
\begin{itemize}
	\item Dynamic Source Routing (DSR) 
	DSR is an on-demand and bandwidth efficient routing protocol, used in multi-hop mobile ad-hoc networks. Each node in DSR protocol must store the complete route to any other node in the network. Moreover, these nodes need to maintain multiple
	routes to a given receiver and this will increase the overhead and make this protocol unsuitable to use for UAVs network with high mobility nodes  \cite{gupta2016survey,johnson1996dynamic}.  
	
	\item Ad-hoc On-demand Distance Vector
	(AODV) is a reactive on demand routing protocol similar to DSR. It is hop-by-hop routing protocol, uses sequence numbers similar to DSDV routing protocol. Therefore AODV is a combination of DSR and DSDV \cite{sahingoz2014networking,murthy1996efficient,perkins2003ad}. 
	In AODV for FANET network with high mobility nodes, the routing table in each node needs to refresh frequently. Therefore, network congestion happened. To solve this problem the authors in \cite{forsmann2007time} proposed a time slotted on-demand routing protocol.
	
	\item Time-Slotted On-demand Routing protocol(TSOR):  
	is a time-slotted version of AODV routing protocol. It is used to minimize the inner node network congestion, where a time component is added to this protocol, as in time-slotted aloha protocol. TSOR mitigates the packet collision and minimizes the network congestion. Moreover, TSOR maximizes the network bandwidth usage, and ensures the delivery of data packets  \cite{bekmezci2013flying,forsmann2007time}.   
	
\end{itemize}

\subsection*{\textbf{5.3.4. Hybrid Routing Protocols (HRP)}}
HRP is a combination of reactive and proactive protocols, that is suitable to be used in large networks. This protocol is proposed to overcome the limitation of reactive and proactive protocols, namely, (1) the large latency of the discovery process can be reduced for reactive protocol, and (2) the overhead of route maintenance occurred by proactive routing protocols can also be reduced. Network in HRP is divided into a number of zones, namely, intra-zone and inter-zone. More specifically, intra-zone routing uses the proactive routing protocol, while inter-zone routing uses reactive routing protocol \cite{gupta2016survey}. 
\begin{itemize}
	\item Zone Routing Protocol (ZRP): is a hybrid routing protocol. ZRP network is divided into a number of zones, each routing zone contains a set of nodes within a predefined zone radius, R \cite{haas2002hybrid,haas2002zone}. In ZPR, there is a tradeoff between the complexity and the improvement of efficiency. This protocol can improve the efficiency of a globally reactive route and can also improve the quality of discovered routes in proactive routing, at the cost of an increase of complexity.
	
	\item Temporarily Ordered Routing Algorithm (TORA): is a hybrid distributed routing protocol that can be used in multi-hop networks. It mainly uses reactive routing protocol, and some of proactive approaches can also be used for protocol enhancement  \cite{park2001temporally}.
\end{itemize}

\subsection*{\textbf{5.3.5. Geographic Based protocol}}
FANETs are characterized by the high mobility of nodes, resulting frequent changes in network topology. Therefore, this will increase the overhead for route maintenance when proactive routing protocol is used. Whilst, the latency for repetitive route discovery process will increase when reactive protocol is used. Location-based routing protocol can be used to solve these problems and to satisfy the FANETs requirements. For example, the routing strategy takes place based on the nodes location, so this protocol assumes that the sender knows the location of the receiver node and the packets can be sent to the coordinates of the receiver node. Consequently, there is no need to maintain the routing information of the network \cite{bekmezci2013flying}.    
\begin{itemize}
	\item Greedy Perimeter Stateless Routing (GPSR):
	is a position based routing protocol. This protocol forwards a packet to the neighbor node geographically closest to the receiver node based on a greedy heuristic. Its uses a beaconing technique to share nodes location information with one-hop neighbors 		
	\cite{shirani2011performance,karp2000gpsr}.
	\item Geographic Position Mobility Oriented Routing (GPMOR): is geographic based protocol suitable to use in FANETs. The conventional position-based routing protocol shares only nodes location information to the best next-hop. The GPMOR can predicts the nodes position using Gaussian-Markov mobility model. Then, this information is used to guess next-hop more accurately.
	\cite{lin2012novel}.
	\item Distance Routing Effect Algorithm for Mobility (DREAM): is a position based routing protocol. In this protocol every node knows its 3D coordinates using GPS and this coordinates is stored into location routing table which will be exchanged with other nodes periodically \cite{peters2011geographical}.
\end{itemize}

\subsection*{\textbf{5.3.6. Hierarchical Routing Protocols}}	
Hierarchical routing protocols are developed to solve the scalability problem of UAVs networks. It can be used when the mission area is large. FANETs are divided into number of clusters, where each cluster consists of cluster head (CH) and cluster members nodes. Here, the CH is within transmission range with all cluster nodes. Moreover, CH is responsible to communicate with upper layers such as satellites and UAVs. CH can also broadcast data to its cluster members such as other UAVs within the cluster. In hierarchical routing, two main routing protocols are used, which are mobility prediction clustering, and clustering algorithm of UAV networking.
\begin{itemize}
	\item Mobility prediction clustering algorithm (MPCA): is a new mobility prediction algorithm based on weighted clustering algorithm to be used in FANETs \cite{zang2011mobility}. 
	MPCA uses dictionary Trie structure prediction algorithm, and link termination time mobility model to predict the network topology updates \cite{konstantopoulos2006mobility}. 
	
	\item Clustering algorithm of UAV networking: This protocol is also proposed to address the frequently changes in FANETs topology. In this protocol, the initial clusters for FANETs is constructed by ground stations. Then the UAVs nodes adjust the clusters based on the change of its locations during the mission \cite{liu2008clustering}.
\end{itemize}

\subsection{Transport Layer}

Due to fluid topology, disconnections and rapid changes in links between UAVs in the FANETs, traditional transport layer protocols such as User Datagram Protocol (UDP) and Transmission Control Protocol (TCP) with flow and congestion control mechanisms cannot be directly implemented for FANETs \cite{fu2005impact}. Despite of its unsuitability, TCP is still used in FANETs. This is because many application layer protocols such as HTTP, FTP depend on TCP.
TCP is a connection-oriented transport protocol, which provides reliability and guarantees ordered delivery of data packets. Moreover, TCP reliability mechanism is based on end-to-end re-transmission as this reduces the network throughput, consumes more energy and increase the transmission delay. On the other hand, UDP is an unreliable transport protocol and it uses a simple connectionless transmission model which does not guarantee packets delivery. 

For FANETs, long delay for packets delivery usually occurred, and the losses of data packets are not caused by congestion, the losses are usually caused by disconnections, rapid changes in links between UAVs and transmissions error occur due to the UAVs mobility \cite{loo2016mobile}. 

Although TCP will provide packets delivery, but the network throughput is not utilized efficiently, any packets loss will be assumed as a congestion. TCP starts with a small amount of packets and probes the network for additional unused link bandwidth. Then, TCP will double the amount of packets each round trip time (RTT) until network congestion occurs (packet loss). When congestion occurs TCP will cut packets amount to half and then begins again network probing by increasing one packet per RTT. This is efficient algorithm for low delays terrestrial networks. But in long delay network such as FANET, this protocol does not utilize the maximum link bandwidth. 

Therefore, the design of congestion control algorithm becomes an important issue for this network, which focuses on modifying and improving this protocol instead of developing new transport protocol for FANETs.  \cite{ivancic2012evaluation}.

Space Communication Protocol Standard - Transport Protocol (SCPS-TP), Satellite Transport Protocol (STP), eXtended STP (XSTP), TCP Peach, TP-Planet, and TCP Westwood (TCPW) are an extension and modification of the TCP implementation to address performance issues of TCP such as long delay, bit error rate and to provide a reliable connection for packets transmissions in space environment \cite{wang2009protocols}.

The Delay and Disruption Tolerant Networking (DTN) architecture was designed to handle the long-delay links, frequent connections interruption environment and in the dynamics environment (high node movements) \cite{shakhatreh2018unmanned}.

\subsection{Application Layer}

Nowadays, the application layer carries most of the internet traffic by HTTP over transport layer protocols TCP and UDP \cite{zanella2014internet}. 
The application layer has the purpose to enable software/user to get access to the networks through transport layer protocols. For example, the transmission of http protocol accomplished by TCP/UDP transport layer protocols and then get access to the network through network layer. The Constrained Application Protocol (CoAP) (RFC 7252) \cite{shelby2014constrained} and Message Queuing Telemetry Transport (MQTT) \cite{banks2014mqtt} are two application layer protocols that can be used for Internet of Things (IoT) and Machine to Machine (M2M) in constrained networks such as FANETs.

CoAP is a UDP based and Representational State Transfer (REST) based web transfer protocol. It is optimized for M2M and IoT applications. It is also designed for constrained devices and networks. CoAP is lightweight, connectionless communication, requires low overhead and low power consumptions and does not provide congestion and flow control \cite{bacco2017application}. On the other hand, MQTT is a TCP based M2M and IoT application protocol. The study in \cite{de2013comparison} provided a comparison between CoAP and MQTT. MQTT and CoAP were envisioned to be the future protocols to connect devices with low power and low bandwidth such as UAVs in FANETs \cite{bruns2017development}.

The research in \cite{scilimati2017industrial} proposed to use UAVs and IoT devices for interacting with the environment to gather data from sensors. These devices have been used with Robot Operating System (ROS) and they use CoAP protocol to interact directly with the UAVs and IoT devices.

The authors in \cite{bruns2017development}  developed a low-cost airborne measurements using UAV platform for geo referencing plots and upcoming observations. In this framework, they used MQTT messaging protocol with rewritten Android APP to measure the sensors data.

\begin{figure}[!h]
	\centering
	\includegraphics[scale=0.425]{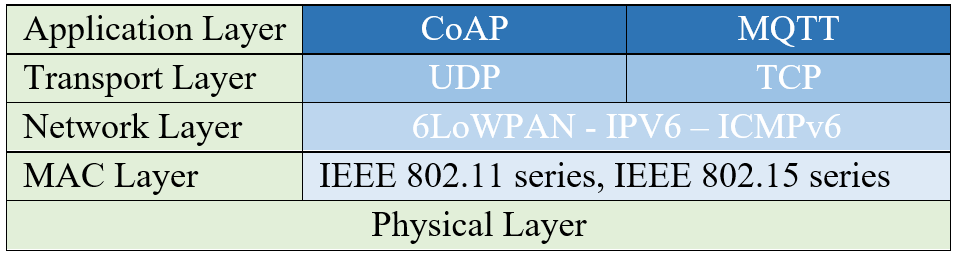}
	\caption{MQTT and CoAP protocol stacks}
	\label{fig:ahmad12}
\end{figure}

Figure~\ref{fig:ahmad12} shows the protocol stacks for MQTT and CoAP protocols. 

\section{Discussion}
\label{discuss}

In this section, we highlight the review findings as listed below:

\textit{UAV deployment challenges}. There are several challenges in the deployment of UAVs as presented in Section~\ref{chall}. Two main challenges are the energy limitations and legislation. For an UAV that uses a single power source, about 80\% of it is used for ascending and hovering. Thus, battery technology development is important. Furthermore, any techniques developed for UAV autonomous operations must take into consideration of the power consumption limitation. Whilst, the UAV-related legislation is necessary and important in civilian applications, to ensure safety for the local and global airspace system.

\textit{Data link classification}. Data link is an important part of UAV communications. From the reviewed literature, we classify the data link based on its operating frequency as presented in Table~\ref{freq}, which includes the advantages and disadvantages of each class.

\textit{Vision-based UAVs collision avoidance approach}. The growth of research activities in vision-based approach, mainly due to the advances in technology to design small and low power sensors and cameras that can be mounted on UAVs. However, the task complexity is one of the challenges that must be addressed due to energy limitation.

\textit{Indoor collision avoidance techniques classification}. From the reviewed literature, we classify the main indoor collision avoidance approaches based on technology and types of processing as presented in Figure~\ref{fig:ahmad6}. This classification is beneficial for future research in this area.

\textit{FANETs networking architecture}. Similarly, the implementation of routing protocol for UAVs network must also take into consideration of the energy limitation as discussed in Section~\ref{Energy}. Based on the reviewed literature, we make a comparison between MANET and FANET as presented in Table~\ref{table}. Furthermore, we reviewed literature on routing protocols based on the OSI communication layer, as presented in Section~\ref{FANET}. We categorize FANET routing protocols in OSI network layer, and provide examples for each protocol.

\section{Conclusion}
\label{con}

In this review paper, we present the general challenges, energy consumption and legislation for deployment of UAVs. Moreover, the main communication services of UAVs have been discussed. Furthermore, we survey the main UAVs collision avoidance approaches and the UAVs network architecture. 

In conclusion, more research are needed to improve collision avoidance approaches in order to find collision free trajectories, and to calculate optimized paths and to develope algorithms for autonomous hovering. Furthermore, more studies are required to develop specific communication protocols for FANETs based on the new networking trends such as Software-Defined Networking (SDN) and DTN.

\bibliographystyle{IEEEtran}
\bibliography{01-paper_preprint}



\end{document}